# IQRray, a new method for Affymetrix microarray quality control, and the homologous organ conservation score, a new benchmark method for quality control metrics

Marta Rosikiewicz[1,2][*], Marc Robinson-Rechavi[1,2]

[1]Department of Ecology and Evolution, University of Lausanne, 1015 Lausanne
[2]Swiss Institute of Bioinformatics, 1015 Lausanne, Switzerland

**ABSTRACT**
**Motivation:** Microarray results accumulated in public repositories are widely re-used in meta-analytical studies and secondary databases. The quality of the data obtained with this technology varies from experiment to experiment and efficient method for quality assessment is necessary to ensure their reliability.
**Results:** The lack of a good benchmark has hampered evaluation of existing methods for quality control. In this study we propose a new independent quality metric that is based on evolutionary conservation of expression profiles. We show, using 11 large organ-specific datasets, that IQRray, a new quality metrics developed by us, exhibits the highest correlation with this reference metric, among 14 metrics tested. IQRray outperforms other methods in identification of poor quality arrays in dataset composed of arrays from many independent experiments. In contrast, the performance of methods designed for detecting outliers in a single experiment like NUSE and RLE was low because of the inability of these method to detect datasets containing only low quality arrays, and the fact that the scores cannot be directly compared between experiments.
**Availability:** The R implementation of IQRray is available at: ftp://lausanne.isb-sib.ch/pub/databases/Bgee/general/IQRray.R

## 1 INTRODUCTION

Thousands of microarray results are available in public repositories such as the Gene Expression Omnibus (Edgar, et al., 2002) and ArrayExpress (Brazma, et al., 2003). This wealth of expression data covering many organisms, tissues, developmental stages, diseases and treatments is now available for meta-analyses, system biology studies, and use in secondary databases. Combining results from several independent studies allows improved detection of differentially expressed genes, and analysis of biological pathways and of co-expression networks (Tseng, et al., 2012). These vast transcriptomic resources have been also extensively used for functional gene annotation and re-analysis of lists of candidate genes obtained with high-throughput experiments. These tasks are facilitated by large secondary databases such as Genevestigator (Hruz, et al., 2008), BioGPS (Wu, et al., 2013), the Gene Expression Atlas (Kapushesky, et al., 2010) or Bgee (Bastian, et al., 2008) that allow mining of many microarray experiments at the same time. Additionally there are many more specialized databases, which for example collect data only from a selected species (Dash, et al., 2012; Le Crom, et al., 2002) or for diseases (Hebestreit, et al., 2012; Rhodes, et al., 2007), or provide resources for more specific analyses, such as COXPRESdb for studying co-expressed genes (Obayashi, et al., 2013) or TiSGeD for the analysis of tissue-specific gene expression (Xiao, et al., 2010).

The quality of results that can be derived from meta-analysis and database searches of microarray data depends directly of the quality of the arrays themselves. Despite this, there are still no objective criteria that allow discrimination between low and high quality arrays (Brettschneider, et al., 2008; Wilkes, et al., 2007). By *quality* in this context we understand the agreement between microarray gene expression level estimations and true gene expression profile. A proper quality control procedure should report whether the amount of biological information captured by the microarray experiment is sufficient to answer a particular biological question.

The most common approach to quality assessment of microarray results makes the assumption that the majority of arrays in each dataset is of good quality, and various parameters serve to identify outlier arrays (Beisvag, et al., 2011; Bolstad, et al., 2005). This task is relatively easy when the dataset is large, but the analysis often lacks power when only a few arrays are analyzed together. Moreover, in the case of a dataset with a high proportion of low quality arrays (for example microarray results derived from degraded RNA samples), the good quality arrays might actually be removed to decrease variability of the dataset. This approach is also powerless to detect datasets composed only of low quality samples.

Most of the existing methods of quality control of Affymetrix microarrays utilize specific features of these chips, such as control probes, pairs of mismatch and perfect match probes, or the fact that there are many probes gathered in probe sets targeting the same transcript. The degradation of transcripts starts from the 3' end, thus the ratio of abundance of 5' to 3' ends serves in molecular biology practice as an indicator of RNA quality. Affymetrix chips contain special probe sets designed to bind 5' or 3' ends of actin and GAPDH transcripts, and the ratio of their signal level is used as a quality parameter (Affymetrix, 2001). Another method that measures RNA degradation takes advantage of the fact that the expression level of each gene is estimated using multiple probes. When probes in every probe set are ordered according to the localization of their binding site in target transcript, the average signal of probes with binding sites closer to the 5' end of transcripts is shifted towards lower values. The slope of the line from the graph illustrating this trend can be used as a quality measure (Gautier, et al., 2004). The most widely used Affymetrix arrays have, for each probe that perfectly matches the target transcript, also a "mismatch probe" that has a single nucleotide mismatch in the middle of the sequence. The mismatch probes were intended to measure the level of unspecific background signal for corresponding perfect match probes. The difference in signal level between pairs of perfect and mismatch probes from the same probe sets is used to generate present/absent calls with the MAS5 algorithm, and the percentage of present calls is one of the quality metrics proposed by the Affymetrix company (Affymetrix, 2001; Wilson and Miller, 2005). Kauffmann, et al. (2009) suggested the inspection of the difference between distributions of signal levels of perfect and mismatch probes as part of quality assessment of microarray experiments. In the current study we quantify this tendency by computing the value of the paired t-test statistic from the signal levels of all perfect/mismatch probe pairs. The Average Background, one of the measures of quality which can be obtained from the original Affymetrix GCOS software, is simply the average of the 2% lowest cell intensities on the chip, and higher values of this parameter suggest high levels of nonspecific binding. The scaling factor is a value that should be used to multiply all values of intensities on the chip in order to scale the 2% trimmed mean signal to a selected constant (Affymetrix, 2001). Finally, the most popular multi-array quality metrics are RLE (Relative Log Expression) and NUSE (Normalized Unscaled Standard Error), which are based on comparisons of the outcome of Robust Multichip Analysis/Probe Level Model (PLM) fitting procedures between arrays from the same experiment (Bolstad, et al., 2004; Gautier, et al., 2004; Irizarry, et al., 2003). In the RLE method medians of probe set expression values are subtracted from expression values of all arrays in experiment. If the quality of a given array doesn't differ greatly from the average quality in the dataset, then such subtracted expression values center around zero and display inter quantile range similar to other arrays. The NUSE values measure precision of estimation of expression values. Shifted or wider distribution of NUSE values indicates problems with the quality of a particular array. The more general version of this parameter, GNUSE (Global Normalized Uncalled Standard Error), compares the standard error of expression estimation to the values stored in pre-computed frozen parameter vectors obtained on the basis of analysis of many arrays of the same type together (McCall, et al., 2011).

---
[*]To whom correspondence should be addressed.

The beneficial influence of removing outlier arrays for microarray data analysis has been demonstrated in several studies (Asare, et al., 2009; Kauffmann and Huber, 2010; McCall, et al., 2011). Development and testing of methods for evaluating the quality of microarray data between experiments suffers from the lack of a good benchmark. In the current study we propose to use the degree of conservation of expression profile between species as an independent indicator of quality, and assess performance of the most popular quality control methods along with a new method developed by us. We show that our new IQRray method is consistently the best in predicting the quality of microarrays.

## 2 MATERIALS AND METHODS

### 2.1 Distribution of probe set average ranks

The IQRray statistic is obtained by ranking all the probes intensities from a given array, and by computing the average rank for each probe set. The Inter Quartile Range of the probe sets average ranks serves then as quality score. The simulated distribution of average ranks from probe sets composed from random probes was obtained by random assignment of ranks to probes from arrays. The simulated distribution of average ranks from probe sets with consistent ranks within probe sets were obtained by assigning succeeding ranks to probes from the same probe sets. In both cases the number of probe sets and the number of probes were identical to arrays of the HG-U133_Plus_2 type. Real examples of low (GSM50702) and high (GSM371402) quality arrays of HG-U133_Plus_2 type were selected from the Bgee database.

### 2.2 Expression data

We selected from the Bgee database (Bastian, et al., 2008) 7 homologous human and mouse organs represented by high numbers of microarray results, obtained using Affymetrix platforms HG-U133_Plus_2 and Mouse430_2 respectively. The number of samples and experiments for each organ are presented in Table 1. All results from prenatal development stages were excluded from the analysis. In the case of the testis dataset only data from adult developmental stage were included. 5 human and 6 mouse datasets composed of arrays from at least 5 independent experiments were used as a training set for benchmarking quality control metrics. The datasets that have not passed this criterion were used as reference only (see Table 1). The complete list of arrays and experiments used in the study is included in supplementary table 1. The initial source of all experiments whose names start with "GSE" is the GEO database (Edgar, et al., 2002), all the other experiments were originally downloaded from ArrayExpress (Brazma, et al., 2003).

Table 1 Number of samples and experiments in organ-specific datasets

| organ name | mouse sample | mouse exp | human sample | human exp |
|---|---|---|---|---|
| blood | 28* | 4* | 429 | 19 |
| liver | 389 | 60 | 51 | 9 |
| kidney | 95 | 15 | 41 | 5 |
| colon | 47 | 6 | 103 | 7 |
| testis | 47 | 11 | 12* | 2* |
| placenta | 50 | 10 | 50 | 9 |
| cerebral cortex | 95 | 11 | 19* | 2* |

*used as reference only.

### 2.3 Computing quality control parameters

The raw data from CEL files were read into the R environment using the package affy (Gautier, et al., 2004) from Bioconductor (Gentleman, et al., 2004). The parameters: average background, percent present, scaling factor and ratios between probe sets for 3' and 5' end of actin and GAPDH transcripts were calculated using the R package simpleaffy (Wilson and Miller, 2005). The slope for RNA degradation was obtained using the package affy (Gautier, et al., 2004). RLE and NUSE metrics were computed with the package affyPLM (Bolstad, et al., 2005) using all arrays that belong to certain experiment. The GNUSE values for every array were computed with the package frma (McCall, et al., 2010) and relevant packages with frma vectors available in Bioconductor. Quality parameters for all arrays used as a training set are included in supplementary table 1.

### 2.4 Preprocessing of raw data

The raw data from CEL files were read into R. The signal values for perfect match (PM) probes were averaged for every probe set. The mapping between probe set IDs and gene IDs were taken from Ensembl version 69 (Flicek, et al., 2013). Probe sets that match more than one gene were excluded. The expression values of independent probe sets that match the same gene were averaged. Only results for genes with one to one orthology between mouse and human with matching probe sets on both microarray types (according to Ensembl 69) were used in further analysis (13136 genes in total).

### 2.5 Correlation of the expression profiles between homologous organs

The pairwise Spearman correlation coefficients between expression profiles of all arrays from organ-specific datasets and homologous organ reference datasets were computed. For every array from an organ-specific dataset the highest correlation coefficient (HOC, the homology organ correlation) obtained was selected and used in further analysis. We used the highest correlation to decrease the probability that the array would be classified as low quality because of natural biological variability related for example to age or sex. The HOC score for all arrays used as a training set are included in supplementary table 1.

### 2.6 Performance of quality metrics

The correlations between quality parameters and HOC scores were calculated using the Spearman correlation coefficient. For some parameters, larger values are better (e.g., percent present), whereas for others smaller values are better (e.g., GNUSE); these are noted as "ascending" or "descending", respectively, in supplementary tables 2a and 2b. Quality scores from all "descending" methods were multiplied by -1 before computing correlations. Correlation for all organs together was obtained after transforming values of HOC score into quartiles, for every organ separately. To evaluate the performance of quality metrics in detecting the lowest quality arrays we selected the worst 5% samples according to each quality control method from all organ-specific datasets, and analyzed the distribution of their quartile of HOC scores. The 5% cut-off value that was used for each quality control method can be found in supplementary tables 2a and 2b.

## 3 RESULTS

### 3.1 IQRray

Because of the limitations of available methods, we propose a new method for multi-experiment quality control. In Affymetrix technology the final expression level is computed on the basis of intensity levels of several independent probes matching the same target mRNA. In our new IQRray method, we transform all probe signal values into ranks, and subsequently compute the average rank of probes that belong to the same probe set. We expect that the higher the quality of a given array, the more consistent the levels of probe signal from the same probe set. The average rank of probes from a probe sets that match highly expressed genes should be high, while the average rank of probes sets that match lowly or not expressed genes should be low. All factors that increase signal noise, such as unspecific hybridization or spatial artifacts, are expected to lead to a more random distribution of probe signals among probe sets. Mixing of low and high ranks in the same probe set should shift the value of the average rank of a probe set towards the average rank of all probes on the array. Consequently, lower quality microarrays will have more narrow spreads of distribution of rank averages. As a measure of this tendency we propose to use Inter Quartile Range (IQR) of probe set average rank: the IQRray score. Figure 1 shows distributions of probe set average ranks from two idealized arrays: one where intensities of probes in probe sets had consistent values and a second where signal values were assigned randomly to the probe sets. We also selected from microarrays in the Bgee database examples of arrays with extreme IQRray scores. It can be seen that the IQR of probe set average ranks is much smaller when the signal values were distributed randomly among probe sets than we they have consistent signal values. The distribution of probe sets average ranks of a presumptive low quality array resembles the distribution of probes with randomly assigned values. The distribution of a presumptive high quality array shows in contrast a bimodal shape due to probe sets targeting lowly or not expressed genes, and highly expressed genes.

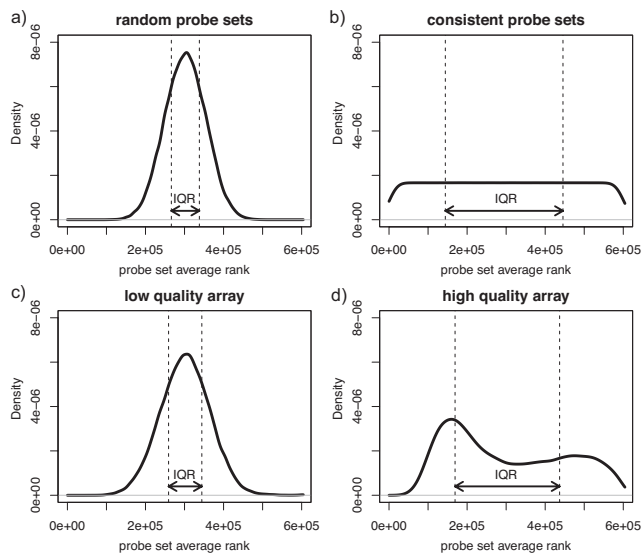

Fig. 1 Distribution of probe set average ranks of a) simulated array with intensity values assigned randomly to probe sets b) simulated array with consistent intensity values in probe sets c) real array with a low IQRray score (GSM50702), d) real array with a high IQRray score (GSM371402)

## 3.2 Benchmarking of QC metrics

In order to compare the performance of the newly proposed method and existing ones for quality control, we set up a test study using 11 large organ-specific datasets – 5 for human and 6 for mice. We use as an external, independent, quality indicator for each microarray the similarity of its expression profile to a reference profile of the homologous organ from the other species. We determined for every array the Homologous Organ Correlation score (HOC score, see Methods). We expect that the lower quality arrays will display less biological signal and consequently a lower correlation with the reference. All arrays analyzed displayed a positive correlation with the profile of homologues organ (fig 2), which is consistent with the previously reported conservation of orthologous gene expression in homologous tissues (Brawand, et al., 2011; Piasecka, et al., 2012; Zheng-Bradley, et al., 2010). However, in each dataset outlier samples of suspicious quality with unusually low HOC scores can be found. These arrays should be preferentially removed through quality control procedure.

## 3.3 Correlation with external quality control method

Each array was evaluated separately by a set of microarray quality control methods. Then we checked how well the quality metrics of each method agreed with the HOC score. There was large variation in the correlation with this external quality indicator (fig 3 and 4). In the case for example of the human blood dataset, the largest dataset in the study (table 1), the IQRray method displayed nearly perfect correlation with the HOC score (Spearman correlation of 0.97) (fig 3a and 4a). In contrast, NUSE and RLE (McCall, et al., 2011), which are frequently used quality control methods, showed only a weak positive correlation (fig 3b and 4b). For this human blood dataset, only a low fraction of samples came from experiments with less than 6 samples (supplementary table1), thus the low correlation with HOC cannot be explained simply by a lack of power, due to an insufficient number of arrays in experiments. In general, across all analyzed datasets NUSE and RLE performed poorly, which suggests that the scores returned by these methods are not directly comparable between independent experiments. All traditional single array quality metrics, such as RNA degradation slope, average background (avbg), scaling factor, and ratios between signal for 3' and 5' end of actin and GAPDH transcripts, show low performance, and the correlation was even negative for some of the methods for some datasets (fig. 4a and b). The score from the GNUSE method, the only published method dedicated to absolute quantification of microarray quality (McCall, et al., 2011), correlates well with the external quality metric only for mouse datasets, while for human GNUSE obtained poor results for nearly all datasets (fig 4b). The IQRray performed the best in 8 out of 11 datasets. The other methods that displayed high agreement with the HOC score for both mouse and human data were percent present and the PM/MM t-test.

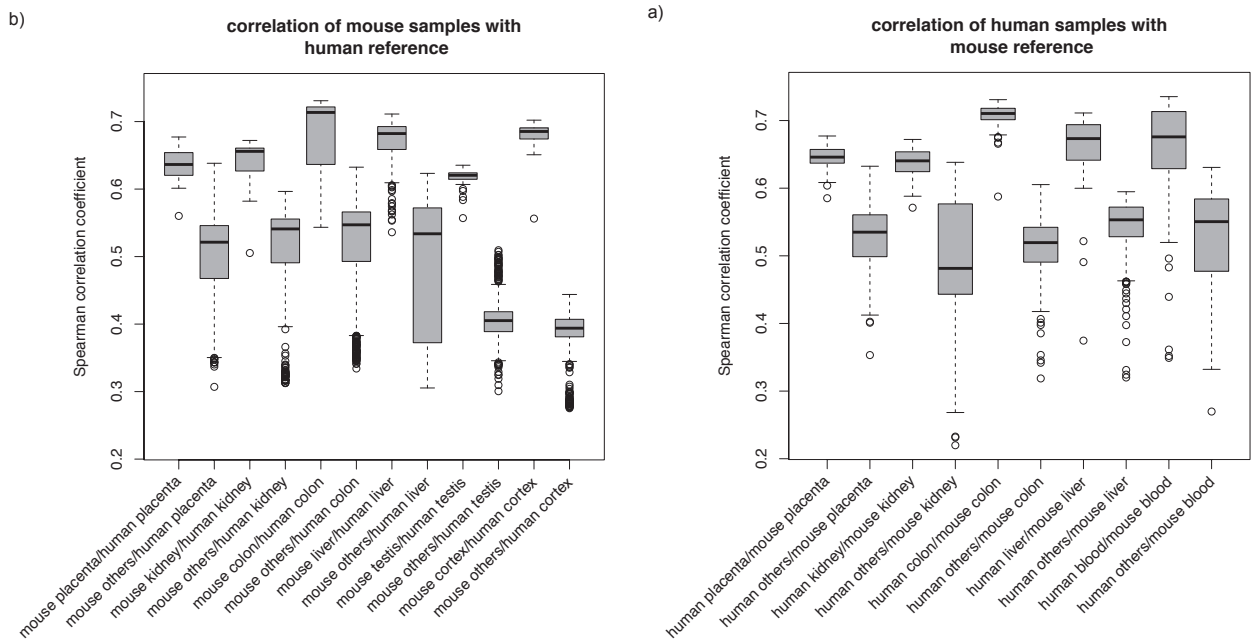

Fig.2 Boxplots of Spearman rho values from correlation test between organ-specific microarray results and reference results from another species a) correlation between human samples and mouse reference b) correlation between mouse samples and human reference

### 3.4 Simulation of quality control assessment of large set of samples.

The ultimate goal of quality assessment is to remove the worst quality samples from datasets. A good quality control method should both correctly identify the worst samples, and avoid assigning falsely low scores to good quality arrays. We simulated a quality control check for a collection of microarrays. We selected the 5% and 10% of samples with the lowest quality according to each quality control method from all samples used in the study (supplementary Table 2). We show the distribution of HOC scores for arrays selected with a 5% cut-off (fig 5). We also measured the efficiency of methods in selecting the arrays with the lowest quality by computing the proportion of results below a chosen cut-off value (supplementary Table 2) displaying a quantile of HOC values also below the same cut-off (e.g., quantiles below 5% for 5% cut-off) (supplementary Table 2). The method that consistently was the most efficient in identifying the worst quality arrays was IQRray. For both mouse and human data all samples selected by this method had quantiles below 0.2, which means that they were of considerably lower quality than other samples derived from the same tissue. This method also identified the highest proportion (around 60% in all cases) of low quality arrays using both 5% and 10% cut-offs. Again, the other methods that performed relatively well were percent present and PM/MM t-test. Consistently with previous results, the GNUSE metric showed high performance for mouse data and confusing results for human data. Among the traditional methods used for quality assessment of microarrays, the scaling factor gave relatively good results for both human and mouse datasets.

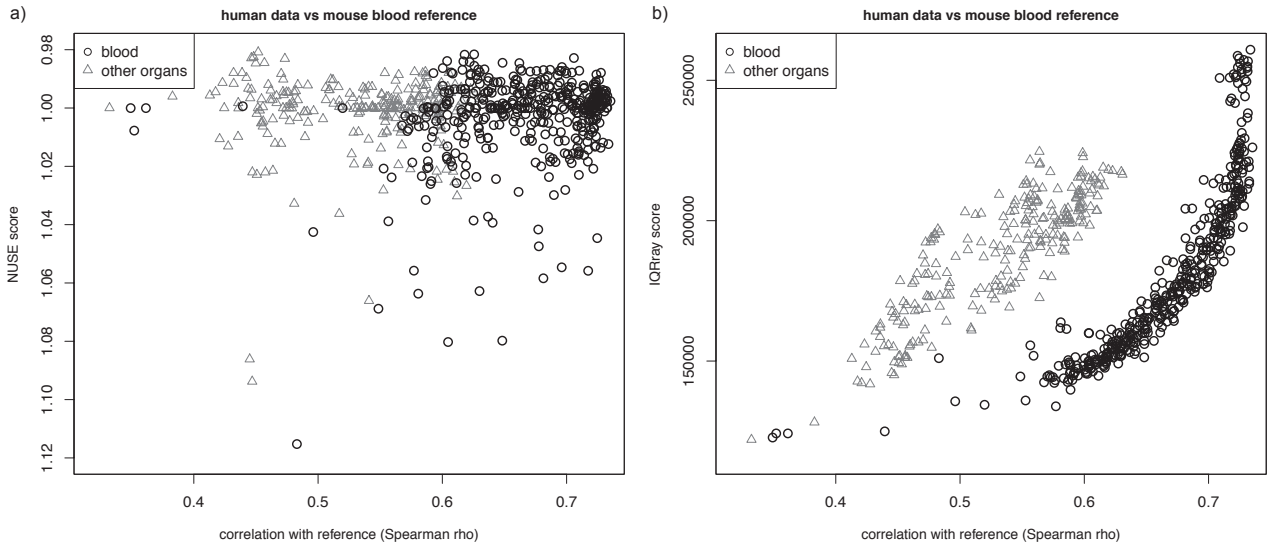

Fig. 3 Correlation between a) NUSE and b) IQRray scores and Spearman rho values from correlation test between mouse blood reference and human blood samples (black circles) and other samples from other organs (grey triangles). Each point on the plot correspond to a single array.

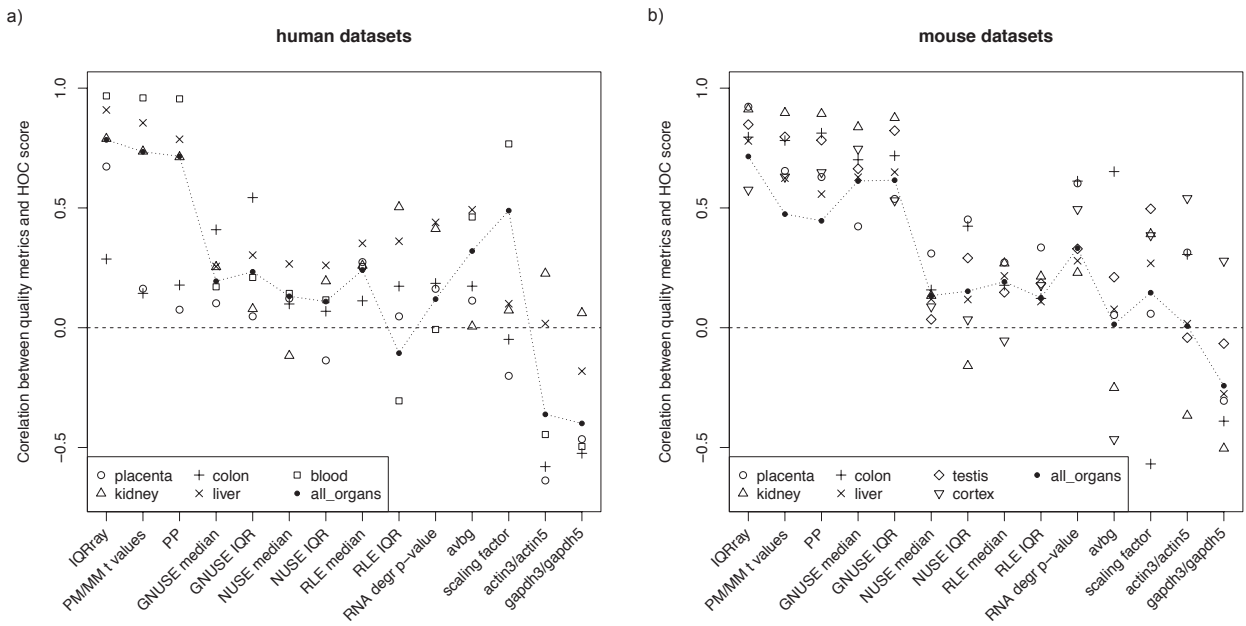

Fig. 4 Spearman rho values from correlation test between quality metrics and HOC score for a) human and b) mouse organ-specific datasets

## DISCUSSION

In this study we introduce a new methodology for benchmarking the quality of results of high-throughput transcriptomic experiments. We decided to use evolutionary conservation of expression profile as an indicator of quality, because it provides an independent assessment of biological relevance. Thanks to the use of correlation between species rather than of correlation between arrays obtained with the same platform, like in previous studies (McCall, et al., 2011) (Gagnon-Bartsch and Speed, 2012), we can avoid giving high scores to low quality results that cluster together. Such spurious results can easily be obtained with microarray technology because the GC content of probes and probe set design, which highly influence the background unspecific level of the measure, remain the same among samples. The HOC can be used not only for the assessment of performance of quality control methods, but more generally for the benchmarking performance of any preprocessing step. Moreover, the method can be directly applied to results from different technologies such as RNA-seq, and may be easily adapted for other types of experiments where an evolutionary conservation of the results is expected, such as ChIP-seq experiments analyzing transcription binding factors or chromatin methylation marks.

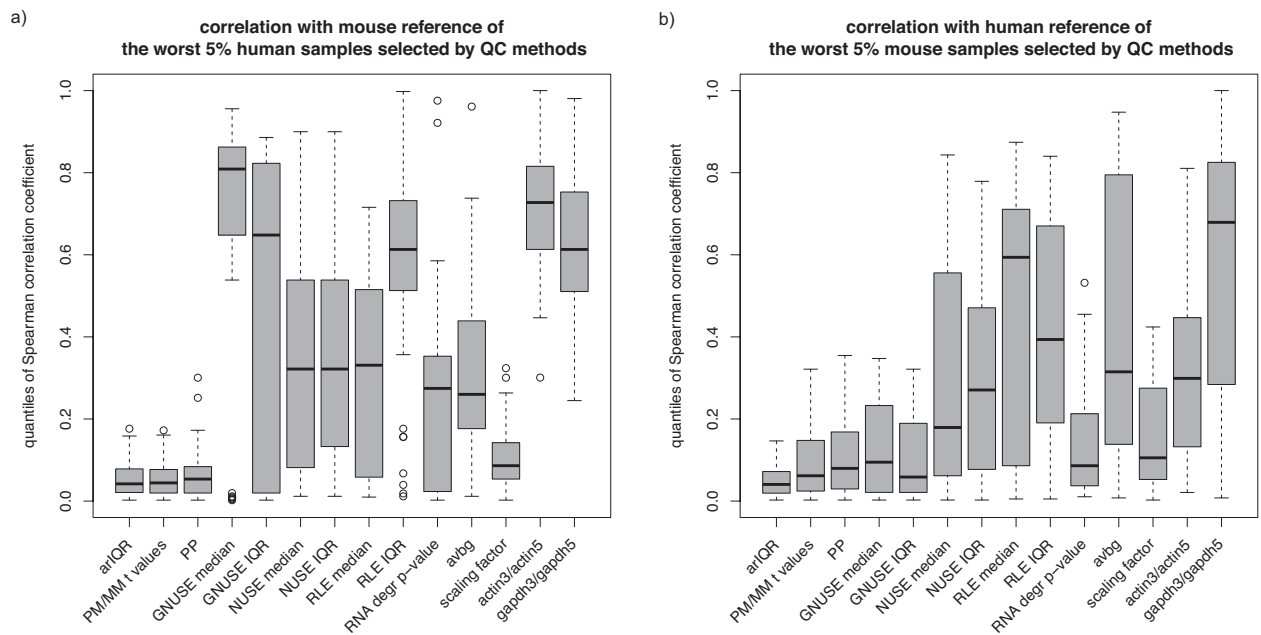

Fig. 5 Boxplot of quantiles of HOC scores of the worst 5% samples selected by different QC methods from a) human b) mouse datasets

A good quality control method should be sensitive to all factors that impede the specificity of binding of labeled targets, such as RNA degradation, insufficient labeling of target transcripts, suboptimal hybridization conditions, or surface defects like bubbles or scratches (Novak, et al., 2002). The IQRray algorithm outperformed all the other tested method in identifying low quality arrays. The method owed its success to two features: 1) the fact that all factors decreasing the strength of specific signal or adding noise to the estimation of the final expression values of a substantial proportion of probe sets will change the distribution of average ranks computed for each probe set; and 2) the possibility of direct comparison of IQRray score between arrays, thanks to the transformation of original values to ranks, which can be considered as a between arrays normalization step. Our study showed that among the known methods for quality assessment of microarrays the best results can be obtained using methodologies based on perfect match and mismatch probe pairs. Use of mismatch probes for estimation of unspecific background signal for perfect match probes has been highly criticized, and gives inferior results in comparison to other method of background signal measurement, mostly because the mismatch probes also bind to some extent the specific target transcript (Irizarry, et al., 2003; Lockhart, et al., 1996; Shedden, et al., 2005). However, when all PM/MM probe pairs on the array are analyzed simultaneously the shift in signal distribution between perfect and mismatch probes is clearly visible, and its strength measured by either PM/MM paired t test or by proportion of probe sets called present seems to be a good indicator of quality. This result might be explained by the fact that the occurrence of differences in hybridization strength between PM and MM probes is strictly dependent on the specificity of target binding, and all factors that negatively influence it also diminish this difference. These methods might be less suitable to detect loss of quality caused by spatial artifacts, although thanks to the fact that the pairs of MM/PM probes are located next to each other on the surface of the microarray the large, intensive surface artifacts also probably have an impact of the final quality score.

The performances of GNUSE scores, the only quality parameter intended for direct comparison of quality between different experiments, differ significantly between mouse and human datasets. Such a discrepancy may indicate that although the assumptions of the methodology are correct, the differences in experimental protocol strongly influence the final results, as was suggested by the GNUSE authors (McCall, et al., 2011). The other possible explanation of this phenomenon is simply lower quality of the frma vector specific for the human HG-U133_Plus_2 array. If a higher proportion of human microarray experiments is of lower quality, then the frozen parameter vector prepared on the basis of these data may underestimate the natural variance in probe signal from the same probe set.

Despite the fact that methods for the analysis of microarray results are already in their maturity, and a lot of effort and attention has been made towards improving methods of quality assessment of microarrays, including calling two international consortia EMERALD and MAQC (Beisvag, et al., 2011; Canales, et al., 2006), there are still no objective rules for defining absolute quality of microarrays. Our new IQRray algorithm for quality control of microarrays appears to be powerful in detecting of low quality arrays, as measured by our independent evolutionary conservation based quality metric.

## ACKNOWLEDGEMENTS

We thank Frédéric Bastian, Aurélie Comte, Nadezda Kryuchkova, Anne Niknejad (UNIL & SIB) for helpful discussions.

*Funding*: Swiss Institute of Bioinformatics (SIB) support to Bgee; Swiss National Science Foundation (grant number 31003A 133011/1); Etat de Vaud.